\documentclass[%
 reprint,
superscriptaddress,
nofootinbib,
 amsmath,amssymb,
 aps,
pra,
a4paper,twocolumn, floatfix
]{revtex4-2}
\usepackage{dsfont}
\usepackage{physics}
\usepackage{graphicx}
\usepackage{dcolumn}
\usepackage{bm}

\usepackage[hidelinks]{hyperref}
\usepackage{caption}
\usepackage[caption=false]{subfig}
\usepackage{ragged2e}
\hypersetup{
    colorlinks,
    linkcolor={red!50!black},
    citecolor={blue!50!black},
    urlcolor={blue!80!black}
}
\usepackage{ragged2e}
\usepackage[]{qcircuit}
\usepackage{mathptmx}
\usepackage{tikz}      
\usetikzlibrary{arrows.meta} 
\usepackage{amsmath}
\usepackage{dsfont}
\usepackage{soul}
\usepackage{tcolorbox} 
\usepackage{xcolor}    
\usepackage{algpseudocode}
\usepackage{comment}
\usepackage{blkarray}

\newcommand{\shiftleft}[2]{\makebox[0pt][r]{\makebox[#1][l]{#2}}}

\tikzset{mycolor/.style = {line width=1bp,color=#1}}%
\tikzset{myfillcolor/.style = {draw,fill=#1}}%

\NewDocumentCommand{\highlight}{O{blue!40} m m}{%
\draw[mycolor=#1] (#2.north west)rectangle (#3.south east);
}
\usepackage{booktabs}
\usepackage{multirow}

\usepackage{mathtools}

\begin{document}

\preprint{APS/123-QED}

\title{Resource-efficient quantum algorithm for linear systems of equations}
\author{Francesco Ghisoni}
 \email{francesco.ghisoni01@universitadipavia.it}
 \affiliation{%
    Dipartimento di Fisica, Università degli Studi di Pavia, via Bassi 6, I-27100 Pavia, Italy
}%

\author{Francesco Scala}
 \affiliation{%
    Dipartimento di Fisica, Università degli Studi di Pavia, via Bassi 6, I-27100 Pavia, Italy
}%

\author{Daniele Bajoni}%
 \affiliation{%
   Dipartimento di Ingegneria Industriale e dell'Informazione, Università degli Studi di Pavia, via Ferrata 1, I-27100 Pavia, Italy
}%

\author{Dario Gerace}
 \affiliation{%
    Dipartimento di Fisica, Università degli Studi di Pavia, via Bassi 6, I-27100 Pavia, Italy
}
%


\date{\today}

\begin{abstract}

Finding the solution to linear systems is at the heart of many applications in science and technology. Over the years a number of algorithms have been proposed to solve this problem on a digital quantum device, yet most of these are too demanding to be applied to the current noisy hardware. In this work, an original algorithmic procedure to solve the Quantum Linear System Problem (QLSP) is presented, which combines ideas from Variational Quantum Algorithms (VQA)
and the framework of classical shadows. The result is the Shadow Quantum
Linear Solver (SQLS), a quantum algorithm solving the QLSP 
avoiding the need
for large controlled unitaries and 
requiring a number of qubits that is logarithmic in the system 
size.
In particular, our heuristics show an exponential advantage of the SQLS in circuit execution per cost function evaluation when compared 
to other notorious variational approaches to solving
linear systems of equations. We test the convergence of the 
SQLS on a number of linear systems, and results 
highlight how the theoretical bounds on the number of resources used by the SQLS are conservative. Finally, we apply this algorithm to a physical problem of practical relevance, by leveraging decomposition theorems from linear algebra to solve the discretized Laplace Equation in a 2D grid.
\end{abstract}
\maketitle
\section{Introduction}\label{sec:intro}
The solution to a linear system of equations 
is at the heart of a wide range of applications in science and 
technology~\cite{WANG2019306, rahola_96, Carpentieri2004753, FRANK2016411}.
In its essence, the problem looks at finding $\Vec{x} \in \mathbb{C}^N$ 
such that $A\Vec{x}=\Vec{b}$, where $A \in \mathbb{C}^{N \times N}$ 
and $\Vec{b} \in \mathbb{C}^N$.  
The computational time for finding the solution is 
affected by: 
the system size $N$, 
the condition number of a matrix $\kappa$, 
the precision error $\epsilon$,
and the sparsity of the matrix. \\
With the first ideas dating back to the 1980's~\cite{Feynman}, 
the field of quantum computing aims at creating 
devices that change the fundamental unit of computation from a 
classical two level system, the bit, to a quantum two level system, 
the qubit~\cite{nielsen00}.
This new paradigm of computation has allowed for the development of 
new quantum algorithms that have a computational
advantage over their classical counterparts~\cite{Shor_1997, grover1996}.

In this context, of particular interest is the work by 
Harrow, Hassadim and Lloyd~\cite{Harrow_2009}, who developed 
a quantum algorithm (also known as the HHL algorithm nowadays) to
solve linear systems of equations. This can be thought as a variation of 
the linear system problem, in which a quantum computer is used to prepare 
a state
$\ket{x} \propto \Vec{x}$, which stores the solution. For this reason, this is also defined as the Quantum Linear System Problem~(QLSP), most often.
Further work following the HHL scheme~\cite{imp1,imp2,imp3,imp4,imp5, 8108698} 
has focused on recasting the HHL algorithm using the Linear Combination of Unitaries (LCU) framework~\cite{Gui_Lu_2006, LongGui-Lu_2008, LongGui-Lu_2009, imp4} allows the creation of an algorithm that 
solves a $N \times N$ 
sparse linear system with, among other features, a logarithmic scaling in $N$, which represents an exponential 
speed up when compared to the linear scaling of the 
best possible classical algorithm~\cite{Harrow_2009}.
It is worth reminding that the exponential advantage of the HHL algorithm 
is at the core of the exponential advantage of many other 
quantum algorithms~\cite{Wiebe_2012, Clader_2013, wang2017efficient, Rebentrost_2014, kerenidis2016quantum}. 
We also note recent work extending quantum linear-system methods to more general matrix equations, such as the Sylvester equation $A\Vec{x} + \Vec{x}B = C$~\cite{somma2025}.
In this setting, the standard linear system $A\Vec{x} = B$ appears as a special case. While
our focus is on the traditional Quantum Linear System Problem (QLSP), these advances
further highlight the growing interest in efficient quantum algorithms for solving
matrix equations of increasing generality.

Despite this, the so called noisy intermediate scale quantum (NISQ) hardware 
available nowadays, limited both in the number of qubits and operation 
reliability~\cite{Preskill_2018}, 
hinders the practical usefulness of any of these algorithms. In fact, the largest linear system of equations solved on quantum hardware with the HHL algorithm 
to date is of size $N=8$ (i.e., corresponding the the Hilbert space size of 3 qubits)~\cite{Wen_2019}. 
This result is due to the
available hardware's inaccuracy in performing large controlled 
operations~\cite{2qubitgates} rather than the available number of qubits, 
which can be in the tens~\cite{PhysRevX.13.041052} 
to hundreds~\cite{Kim2023} of qubits 
depending on the architecture. \\
In the meantime, so called Variational Quantum Algorithms~(VQAs)~\cite{Cerezo_2021} have been developing as a class of hybrid classical-quantum algorithms to make use of the available NISQ hardware in the nearest terms.
These algorithms run a series of parameterized quantum circuits~\cite{PQA1}, 
in which the parameters are iteratively updated by using classical 
optimization techniques. VQAs have been proposed to solve a wide range of 
problems, such as classification~\cite{VQClassifier}, 
chemical simulation~\cite{chem1,chem2,chem3,chem4}
entanglement witness~\cite{Scala_2022}, solving
non-linear differential equations~\cite{Lubasch_2020, pool2024nonlineardynamicsgroundstatesolution}. Alternatives to the HHL algorithm based on VQAs have also been proposed~\cite{Pellow-Jarman2023}.
In particular, the Variational Quantum Linear Solver~(VQLS)~\cite{VQLS} is the most promising variational approach due to its efficient scaling 
in $N$ as well as its best resource usage.
As a result, the VQLS has recently become the focus of numerous applications to real-world linear systems~\cite{Liu_2021_Poisson, LIU2024116494, Liu_2022, LUO2024102070, xing2023, e25040580}.
Nevertheless, in most cases of practical interest this algorithm still requires  
large circuit depths, or large number of qubits, depending on the type of cost function to be evaluated on actual quantum hardware. 

\begin{figure*}[ht]
    \includegraphics[scale=0.57]{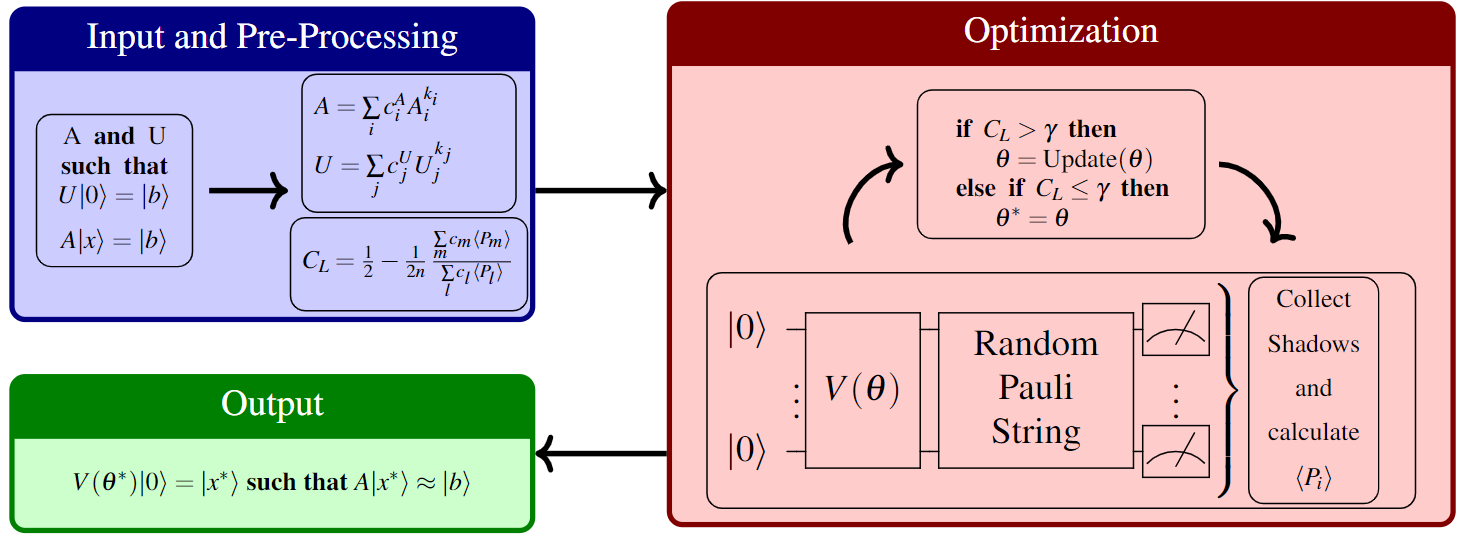}
    \caption{\justifying Schematic diagram for the SQLS. The aim is to solve a linear system $A\Vec{x_0} = \Vec{b}$. The inputs are: a matrix $A$, a unitary $U$, both written as a linear combination of Pauli strings, and form a linear system of Equations, $A\ket{x} = \ket{b}$ in which $U\ket{0} = \ket{b}$. The solution is found by using a hybrid classical-quantum variational algorithm were the parameters $\Vec{\theta}$ of a parameterized quantum circuit, $V(\Vec{\theta})$, will be optimized through classical optimization techniques. Due to the assumptions on the form of $A$ and $U$, the cost function, $C_L$ (Eq.~\eqref{eq:local cost}), involves the calculation of linear sums of expectation values of Pauli strings, which can be calculated using a small number of shallow circuits through classical shadows. The optimization process terminates when the condition $C_L \leq \gamma$ (see Eq.~\eqref{eq:bound}) is reached, returning a set of parameters $\theta^*$ such that $V(\theta^*)\ket{0}=\ket{x^*} \approx \frac{1}{||x_0||} \Vec{x_0}$.}
    \label{fig:SQLS}
\end{figure*}

On a more general ground, the quest for optimal resource usage of quantum circuits is at the core of various protocols in quantum 
information~\cite{D_Ariano_2007,aaronson2018shadow}.
One such framework has been recently proposed as the \textit{classical shadows}~\cite{Huang_2020}, a resource-efficient and information-complete measurement scheme that uses $N_{shadow}$ copies of a 
state $\rho$ to create the so called classical shadow $S(\rho, N_{shadow})$ of that state. Among other applications, the shadow can be used to 
accurately estimate linear functions, in particular. Due to their flexibility, classical shadows
have been employed in combination with VQAs for a wide range of applications, such as warm starting~\cite{truger2024warmstarting}, 
optimizing circuits~\cite{basheer2022alternating}, and to
avoid barren plateaus~\cite{PRXQuantum.3.020365}.

In this work, essentially we merge ideas from VQAs and classical shadows 
to present a procedure to solve the QLSP. Our original procedure, which we define as the Shadow Quantum Linear Solver~(SQLS), leverages classical shadows to efficiently evaluate
the cost function encoding the solution to a QLSP. 
When compared to other variational approaches to the QLSP, we notice that the SQLS uses less qubits, 
less controlled operations, 
and shallower circuits. Furthermore, in the noiseless hardware limit the SQLS offers an efficient scaling with the systems size, $N$, allowing to address  problems that might be difficult to be solved by only using classical computing resources.

The present manuscript is organized as follows:
in Sec.~\ref{sec:theoreticalbackground} we review 
the basics of VQAs  
and the classical shadows formalism, for completeness. 
In Sec.~\ref{sec:sqls} we introduce the SQLS, showing the theoretical advantages
expected in the solution of the QLSP.
In Sec.~\ref{sec:results} we present the studies carried out in this work to characterize the SQLS: First, we theoretically show that SQLS is less resource intensive for 
more complex systems of equations when compared to other variational 
approaches to the QLSP;   
finally, we leverage techniques from linear
algebra to find the solution to a real physics problem, the discretized
Laplace Equation on a grid. 
Ultimately, we believe that the procedure hereby introduced
has the potential to speed up the advent of near-term solutions
of the QLSP on NISQ devices. 
In addition, the paper brings to attention further techniques to transform highly non-trivial linear systems into a QLSP, which could inspire new applications of quantum algorithms to solve linear systems of equations.

\section{Theoretical Background}
\label{sec:theoreticalbackground}
Here we report the theoretical background required to formalize the Shadow Quantum Linear Solver (SQLS). 
In particular, we first summarize Variational Quantum Algorithms (VQAs), and then outline the main aspects of classical
shadows. Throughout the manuscript, we will adopt the convention to express Pauli gates as $\sigma^i$, in which $i \in \{ x,y,z\}$, the $2 \times 2$ identity gate as $\mathbb{I}$ and
$k$-local Pauli strings as $P^k = \bigotimes_{j=1}^{k} \sigma^{j} \otimes \mathbb{I}_k-$, where $\mathbb{I}_k-$ denotes identity on all qubits except the $k$-th qubit. Furthermore, we note that in the whole manuscript we use the shorthand notation
$\log_2(\cdot) \equiv \log(\cdot)$.

\subsection{Variational Quantum Algorithms}
VQAs are a class of hybrid quantum-classical algorithms aimed at distributing the computation between a digital quantum device and 
a classical computer, making them ideal for the currently available
NISQ devices~\cite{Cerezo_2021}. 
These algorithms require the definition of a cost function $C$ (often also called the \textit{loss} function) 
whose minimum represents the optimal solution of the problem. 
Following this, a generic parameterized quantum circuit~\cite{PQA1}
is proposed. This circuit, also known as the \textit{ansatz}, can be expressed as:
\begin{equation}
    \label{ansatz}
    V(\theta) = G_{k_L}(\theta_{L})G_{k_{L-1}}(\theta_{L-1})...G_{k_1}(\theta_{1}),
\end{equation}
in which the quantum gates $G$ are chosen from an alphabet 
$\mathcal{A}=\{ G_k(\theta_i) \}$,
where $\Vec{k}$ dictates the type and order of the gates 
and $\theta_i$ are continuous parameters.
Since the cost function is defined in terms of the anstaz, it is dependent on these parameters as well, 
i.e. $C(\Vec{\theta})$.
The goal is then to find the optimal set of parameters $\Vec{\theta}^*$
by minimizing the cost function,
\begin{equation}
\label{eq:mincost}
    \Vec{\theta}^* = \text{arg} \; \min\limits_{\Vec{\theta}} \left\{C(\Vec{\theta})\right\}.
\end{equation}
In the setting of Equation~\eqref{eq:mincost}, 
$\Vec{\theta}^*$ can be found by using classical optimization 
techniques.
Despite this simple formulation, VQAs may suffer from 
flat optimization landscapes, known as 
\textit{barren plateaus}, which renders both gradient based and gradient
free optimization techniques poorly converging~\cite{McClean_2018}. 
This phenomenon has been studied in depth in the literature, and various 
causes and possible solutions have been identified~\cite{larocca2024reviewbarrenplateausvariational}. 
Among these, we point out two potential solutions: the first calls for the use of a \emph{local} cost function, in which the quantum circuits undergo only single-qubit measurements~\cite{Cerezo_2021_local}, while the second suggests the initialization of the parameters to small values~\cite{zhang2022escapingbarrenplateaugaussian, Park_2024, wang2023trainabilityenhancementparameterizedquantum}. 

A further feature that some VQAs have shown to exhibit is  
Optimal Parameter Resilience~(OPR)~\cite{Sharma_2020}. 
OPR refers to the phenomenon where the optimal parameters
are not affected by a particular type of noise. In particular let
$\Tilde{C}_L$ be the noisy cost function of a VQA, then OPR 
guarantees that $\min \Tilde{C}_L = \min C_L$. 
VQAs which are resilient to a large number of noise models are considered
especially well suited for NISQ devices.\\
\subsection{Classical Shadows}
\label{subsec:classicalshadows}
Classical shadows have been formulated as a protocol allowing to construct 
a minimal representation
$S(\rho;N_{shadow})$ of a quantum state $\rho$ to be employed to accurately estimate functions,
both linear and nonlinear,
with a given error $\epsilon_{shadow}$. 
When applying the procedure, we assume to have access to a fixed but unknown $n$-qubit quantum state 
$\rho$. The protocol is then applied by starting with a unitary $U$ randomly chosen from a fixed ensemble $\mathcal{U}$, and then measured 
in the computational basis.
The outcome, $\ket{\alpha} \in \{\ket{0},\ket{1}\}^{\otimes n}$, is used to calculate
and store an approximation of $\rho$, i.e.,
\begin{equation}
    \hat{\rho} = \mathcal{M}^{-1} \left( U^{\dagger} \dyad{\alpha}{\alpha} U \right),
\end{equation}
where $\mathcal{M}$ is a quantum channel defined as 
\begin{equation}
    \label{eq:channel}
    \mathcal{M}(\hat{\rho}) = \mathbb{E}\left[ U^{\dagger} \dyad{\alpha}{\alpha} U \right].
\end{equation}
If the set of unitaries $\mathcal{U}$ is 
tomographically complete, i.e., for each $\sigma \neq \rho$
there exists $U \in \mathcal{U}$ and $b$ such that 
$\bra{b}U\sigma U^{\dagger}\ket{b} \neq \bra{b}U\rho U^{\dagger}\ket{b}$, then it has been shown 
that $\mathcal{M}$ can be always inverted~\cite{Huang_2020}. Repeating this procedure
$N_{shadow}$ times, and saving each result $\hat{\rho_i}$ in a matrix, finally produces the \emph{classical shadow}, or shadow, 
$S(\rho; N_{shadow})$ defined as: 
\begin{equation}
    \label{eq:classicalshadow}
    S(\rho; N_{shadow}) = \left\{ \hat{\rho}_1, \hat{\rho}_2, ..., \hat{\rho}_{N_{shadow}}\right\} .
\end{equation}
This shadow can be used to estimate a wide variety of quantities,
but for the purpose of the SQLS we only present the algorithm 
and the theoretical guarantees to use $S(\rho; N_{shadow})$ to calculate 
$M$ expectation values  
$\{ \langle P_i^{k_i} \rangle \}$, where 
$P_i^{k_i}$ are $k_i$-local Pauli strings. 
In this case, the quantity  $\langle P_i^{k_i} \rangle$ 
is estimated by computing : 
\begin{equation}
    \label{eq:shadowalgorithm}
    \langle P_i^{k_i} \rangle = \text{median} \left\{ \text{Tr}\left( P_i^{k_i} \hat{\rho}_1 \right), \text{Tr}\left( P_i^{k_i} \hat{\rho}_2 \right),...,\text{Tr}\left( P_i^{k_i} \hat{\rho}_{N_{shadow}} \right) \right\},
\end{equation}
in which $N_{shadow}$ depends on the error $\epsilon_{shadow}$, and scales as
\begin{equation}
    \label{eq:shadowbound}
    N_{shadow} \propto \frac{\log(M)3^{k}}{\epsilon_{\text{shadow}}^2}.
\end{equation}
Furthermore we note that the same amount of shadows can be used 
to compute the density matrix $\rho$ with a error 
$\epsilon_{shadow}$ by computing:
\begin{equation}
    \label{eq:tomography}
    \rho = \mathbb{E}(S(\rho, N_{shadow})).
\end{equation}
\section{Shadow Quantum Linear Solver}
\label{sec:sqls}
Inspired by the variational approaches to the QLSP, we hereby present the Shadow Quantum Linear Solver (SQLS), a new resource efficient VQA aimed at leveraging the power of classical shadows in the context of solving  linear systems of equations. 
Our proposal claims efficient resource utilization as 
well as, in the noiseless limit, 
an efficient scaling with the system size.
A schematic diagram of the proposed algorithm is represented in Fig.~\ref{fig:SQLS}. \\
Considering a $N \times N$  QLSP of the form 
$A\ket{x_0} = \ket{b}$, the SQLS aims at finding a set of 
optimal parameters $\Vec{\theta^*}$  that, when used in a
quantum circuit $V$, they allow to prepare the solution to the QLSP, i.e.,
$V(\Vec{\theta^*})\ket{0} = \ket{x(\Vec{\theta^*})}=\ket{x^*}\approx\ket{x_0}$. 
Hence, the SQLS requires $n=\log(N)$ qubits and two inputs:
1) a unitary $U$ that satisfies: 
$U\ket{0} = \ket{b}$ and, 
2) a decomposition of the matrices $A$ and $U$ into a linear combination of $k$-local Pauli strings $P^{k}$, i.e., 
\begin{align} 
\label{eq:paulidecomp_A}
A & = \sum\limits_{i=1}^{L_A} c_i^A A_i^{k_i} \\
\label{eq:paulidecomp_U}
U & = \sum\limits_{j=1}^{L_U} c_j^U U_j^{k_j},
\end{align}
in which $c_i^A, c_j^U \in \mathbb{C}$, and $A_i^{k_i}$ and $U_j^{k_j}$ 
are $k_i$-local and $k_j$-local Pauli strings, respectively. 
Furthermore, we note that this form of $U$ is equivalent to assuming 
that the operator is given in an efficient gate sequence, whilst the form of 
$A$ is typical of VQAs~\cite{chem1}.
Finally, we assume that: (1) the condition number of the matrix 
$A$ is finite, $\kappa_A < \infty$, 
(2) the $l_2$ norm of $A$ is bound by 1, $\|A\|_2 \leq 1$, (3) 
$L_A$ and $L_U$ are polynomial in $n$,  and (4)
the Pauli matrices $A_i^{k_i}$ and $U_i^{k_j}$ have a low locality. The assumption that $L_A$ and $L_U$ scale polynomially with the number of qubits, $n$, can also be motivated by physically relevant settings. In many applications, $A$ represents a local operator, such as a $k$-local Hamiltonian or a discretized Laplacian on a regular grid.
In these cases, each degree of freedom only couples to a constant number of neighbours,
so $A$ is sparse and it can be decomposed into a sum of $k$-local Pauli strings with
$L_A = \mathrm{poly}(n)$. A concrete example of this is the Laplace operator used in
Sec.~\ref{sec:Laplace}. For $U$, we assume an implementation by a polynomial-depth circuit composed of one- and two-qubit gates, which likewise leads to a decomposition scaling as $L_U = \mathrm{poly}(n)$. The last assumption can be seen as considering a \emph{sparse} linear system, which is common in QLSP
algorithms~\cite{Harrow_2009}. Furthermore, in this regime there exist algorithms 
~\cite{pesce2021h2zixy,Vidal_Romero_2023,bergholm2022pennylane,
jones2024decomposing, hantzko2024tensorized}
that efficiently decompose a sparse unitary as a linear sum of Pauli strings. We note that the SQLS is not an LCU based algorithm since we do not rely on block encoding of the matrix, rather the solution is being stored through an amplitude encoding.

With these inputs, the SQLS runs an 
optimization process using a local cost function analogue to the 
one proposed in Ref.~\cite{VQLS}. 
In particular, given the un-normalized quantum state $\ket{\psi(\Vec{\theta})} = A\ket{x(\Vec{\theta})}$,
the local cost function is defined as:
\begin{equation}
    \label{eq:local cost}
    C_L(\Vec{\theta}) = \frac{\bra{x(\Vec{\theta})}H_L\ket{x(\Vec{\theta})}}{\braket{\psi(\Vec{\theta})}{ \psi(\Vec{\theta})}}=
    \frac{\bra{x}H_L\ket{x}}{\braket{\psi}{ \psi}}, 
\end{equation}
in which we employ the compact notation $\ket{x(\Vec{\theta})}=\ket{x}$, and the effective Hamiltonian is explicitly expressed as:
\begin{equation}
    \label{eq:hamiltonian}
    H_L = A^{\dagger} U \left( \mathbb{I} - \frac{1}{n}\sum\limits_{r=1}^n \dyad{0_r}{0_r} \otimes \mathbb{I}_r-\right) U^{\dagger}A\, .
\end{equation}
Where $\mathbb{I}_r-$ is the identity on all qubits
except for the $j$-th one, $\ket{0_j}$ is the $j$-th qubit zero logical state. The cost function in Eq.~\ref{eq:local cost} effectively encodes the solution to the problem: if $C_L \rightarrow 0$, then $\ket{\psi} \rightarrow \ket{b}$, meaning that a minimization of $C_L$ guarantees a solution to the QLSP. 
We now introduce the approximation error $\epsilon$ as:
\begin{equation}
    \label{eq:error}
    \epsilon = \frac{1}{2}\text{Tr} \left( | \dyad{x_0}{x_0} - \dyad{x^*}{x^*} | \right),
\end{equation}
which is the trace distance between the real solution, 
$\ket{x_0}$, and the approximate optimal solution, 
$\ket{x^*}$. 
Then, considering a target approximation error $\epsilon$, 
for a QLSP with the condition number $\kappa$ and 
requiring a number of qubits $n$, 
the local cost function from Eq.~\eqref{eq:local cost} is bound by
\begin{equation}
    \label{eq:bound}
    C_L \geq \frac{1}{n}\frac{\epsilon^2}{\kappa^2} = \gamma
\end{equation}
in which $\gamma$ is the termination threshold for the optimization process~\cite{VQLS}. 
Furthermore, it has been shown that for a $N \times N$ QLSP with condition number $\kappa$, and accepting a final error $\epsilon$, the noiseless cost function scales as follows~\cite{VQLS}:
poly logarithmically in the problem size $N$, 
linearly in $\kappa$,
and logarithmically in $1/\epsilon$.
It is important to notice that the system size scaling guarantees an exponential advantage of the noiseless SQLS over the best classical 
algorithm~\cite{Harrow_2009}. 
Furthermore, it has been proposed that 
estimating the noiseless cost function to a $1/ \text{poly}(n)$ precision  
belongs to the  
\emph{Determinisitic Quantum Computing with 1 clean qubit} (\textbf{DQC1}) complexity class.
The latter contains all problems that can be efficiently solved
with bounded error in the one-clean-qubit model of
computation~\cite{Knill_1998}, which has been suggested to 
be impossible to simulate on a classical computer~\cite{Morimae_2017, Fujii_2018}.\\
Given the setting of the linear system introduced in Eqs.~\eqref{eq:paulidecomp_A} and~\eqref{eq:paulidecomp_U}, the cost function can be re-written in a compact notation as 
\begin{equation} \label{eq:local_cost_2}
C_L = \frac{1}{2}- \frac{1}{2n}\frac{\mu}{\omega} \, ,
\end{equation}
in which 
\begin{align}
    \label{eq:mu}
\mu &= \sum\limits_{r=1}^{n} 
    \sum\limits_{i,j=1}^{L_A} \sum\limits_{l,p=1}^{L_U}c^A_ic_{j}^{A*}c^U_pc_{l}^{U*} \delta_{ijlp}^{r}  \\
\label{eq:omega}
\omega &= \sum\limits_{i,j=1}^{L_A}c^A_ic_{j}^{A*}\beta_{ij} \, ,
\end{align}
with $\delta_{ijlp}^{r} = \langle x | A_j^{k_j\dagger} U_p^{k_p} \sigma^z_r U^{k_l\dagger}_l A^{k_i}_i |x\rangle$
and~$\beta_{ij} = \langle x| A_j^{k_j\dagger} A_i^{k_i}|x\rangle$.
Here, $\sigma^z_r$ represents a $\sigma^z$ on the $r$-th qubit. The derivation of how these equations come about can be found in App.~\ref{sec:costfuncderivation}.
At this stage, it is important to point out that $\mu$ and $\omega$ are, essentially, weighted sums of expectation values of Pauli strings.
Since the evaluation of the cost function requires the
estimation of many expectation values of Pauli strings, one can leverage 
classical shadows, and in particular Eq.~\eqref{eq:shadowalgorithm}, to efficiently perform this operation. More in detail, given $\ket{x}$ with 
associated density matrix $\rho_x$, we define its classical shadow as 
$S(\rho_x, N_{shadow})$, where $N_{shadow}$
is determined following Eq.~\eqref{eq:shadowbound}.
Due to the very efficient resource use of classical shadows, 
the advantage of using the SQLS is 
two-fold:
circuits are shallow, and
$N_{shadow}$ scales logarithmically with the number of 
expectation values to calculate. A full investigation of the resources
required in a SQLS run is given in  Sec.~\ref{sec:resources}.\\
Once the optimization process is over, Eq.~\eqref{eq:tomography} can also be used to reconstruct the density matrix of the state $\ket{x^*} \approx \ket{x_0}$ without any additional measurement, since classical shadows  also allow for quantum state tomography~\cite{Huang_2020}. 
This is a non-trivial addition to have as part of the algorithm, 
since the state vector reconstruction requires, in general, an exponential
amount of measurements to the circuit~\cite{MatMul}.
For example, both in the HHL and the VQLS the solution to the linear
system can only be extracted through tomographic procedures,
which would cancel out the exponential advantage offered by running both 
algorithms. Therefore, eliminating the need for a final tomography of the output quantum state the SQLS
significantly reduces both the amount of resources required as well as
the computational time.\\
Finally, we address the Optimal Parameter Resilience(OPR) of the SQLS. In fact,
in~\cite{VQLS} it was proven that the cost function of the form of 
Equation~\eqref{eq:local cost}, when evaluated with the hadamard test,
exhibits OPR to both a global depolarizing 
channel as well as measurement noise. 
We argue that the OPR to a global depolarizing
channel is inherited in the SQLS whilst, due to the fact that shadow
collection requires multiple qubit measurements, OPR to measuement noise
is not inherited. The qualitative arguments for OPR in the SQLS can be found in 
App.~\ref{sec:OPR_res}.\\ 
To summarize the SQLS algorithm, where the main steps are reported in Fig.~\ref{fig:SQLS}, given the inputs of 
Eqs.~\eqref{eq:paulidecomp_A} and~\eqref{eq:paulidecomp_U}. 
First, we define a parametrized ansatz $V$, and initialize it with a set of parameters
$\Vec{\theta}$. This is followed by the optimization process, 
which involves the creation of the classical shadow of the ansatz state $V(\Vec{\theta})\ket{0}$, followed
by the calculation of the local cost function $C_L$ as defined in  
Eq.~\eqref{eq:local_cost_2}, and the update of parameters through
classical optimization techniques.
This optimization step is repeated until the termination threshold
$\gamma$ (defined in Eq.~\eqref{eq:bound}) is met by $C_L$. This will terminate the 
optimization process and the SQLS will give the state $\ket{x^*}$ as its output, 
i.e., the state in which the solution to the QLSP is encoded.

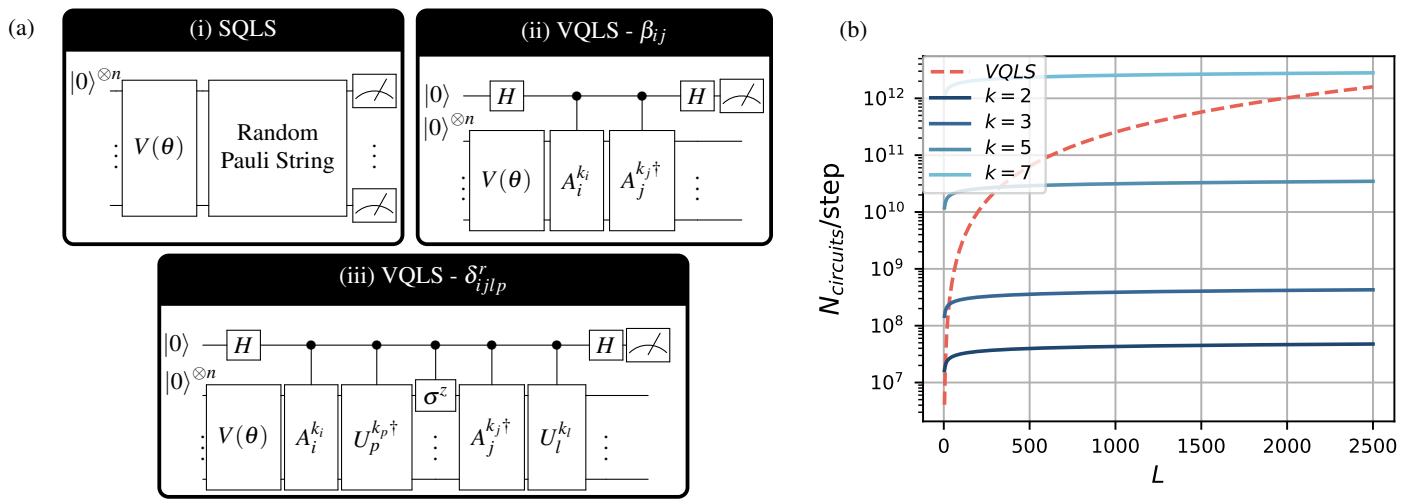
\begin{figure*}[!ht] 
    \centering
    \begin{minipage}[t]{0.54\textwidth}
        \centering
            \begin{tikzpicture}
        \node at (-25,0) {
        \begin{tcolorbox}
        [colback=white,colframe=black,title= (i) SQLS, width=4.5cm, height=3.15cm, center title]
        \Qcircuit @C=0.25em @R=1.25em {
            &\lstick{} & \qw & \multigate{2}{V(\theta)} & \qw & \multigate{2}{\parbox{1.5cm}{\centering Random\\ Pauli String}} & \meter\\
            && \vdots  & \nghost{V(\theta)} && \nghost{\parbox{1.5cm}{\centering Random\\ Pauli String}} & \vdots\\
            &\lstick{} & \qw & \ghost{V(\theta)} & \qw & \ghost{\parbox{1.5cm}{\centering Random\\ Pauli String}} & \meter 
            \inputgrouph{1}{3}{-0.4em}{\ket{0}^{\otimes n}}{0.35em}
            }
        \end{tcolorbox}};

        \node at (-22.5,-3.3) {
        \begin{tcolorbox}
        [colback=white,colframe=black,title= (iii) VQLS - $\delta_{ijlp}^r$, width=7.cm, height=3.25cm, center title]
        \Qcircuit @C=0.15em @R=0.75em {
    &\lstick{\ket{0}} & \gate{H} & \ctrl{1} & \ctrl{1} & \ctrl{1} & \ctrl{1} & \ctrl{1} & \gate{H} & \meter\\
    &\lstick{} & \multigate{2}{V(\theta)} & \multigate{2}{A^{k_i}_i} & \multigate{2}{U^{k_p\dagger}_p} & \gate{\sigma^z} & \multigate{2}{A^{k_j\dagger}_j} & \multigate{2}{U^{k_l}_l} & \qw & \qw\\
    &\vdots & \nghost{V(\theta)} & \nghost{A^{k_i}_i} & \nghost{U^{k_p\dagger}_p} & \vdots & \nghost{A^{k_j\dagger}_j} & \nghost{U^{k_l}_l} & \vdots \\
    & \lstick{} & \ghost{V(\theta)} & \ghost{A^{k_i}_i} & \ghost{U^{k_p\dagger}_p} & \qw & \ghost{A^{k_j\dagger}_j} & \ghost{U^{k_l}_l} & \qw & \qw
    \inputgrouph{2}{4}{-0.6em}{\ket{0}^{\otimes n}}{0.35em}
            } 
        \end{tcolorbox}};

    \node at (-20.25,0) {
        \begin{tcolorbox}
        [colback=white,colframe=black,title= (ii) VQLS - $\beta_{ij}$, width=4.7cm, height=3.15cm, center title]
        \Qcircuit @C=0.25em @R=0.75em {
    & \lstick{\ket{0}} & \gate{H} & \ctrl{1} & \ctrl{1} & \gate{H} & \meter\\
    & \lstick{} & \multigate{2}{V(\theta)} & \multigate{2}{A^{k_i}_i} & \multigate{2}{A^{k_j\dagger}_j}  & \qw & \qw\\
    & \vdots & \nghost{V(\theta)} & \nghost{A^{k_i}_i} & \nghost{A^{k_j\dagger}_j} & \vdots \\
    & \lstick{} & \ghost{V(\theta)} & \ghost{A^{k_i}_i} & \ghost{A^{k_j\dagger}_j} & \qw & \qw
    \inputgrouph{2}{4}{-0.6em}{\ket{0}^{\otimes n}}{0.35em}
            } 
        \end{tcolorbox}};
    \end{tikzpicture}
    \shiftleft{10.25cm}{\raisebox{6.25cm}[0cm][0cm]{(a)}}
    \end{minipage}
    \hspace{0.001\textwidth}
    \begin{minipage}[t]{0.45\textwidth}
        \centering
        \includegraphics[width=\linewidth]{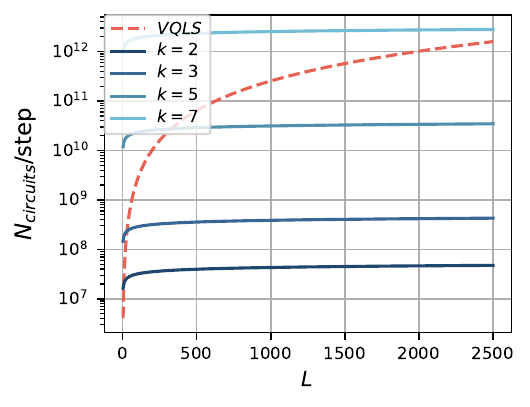}
        \shiftleft{3.5cm}{\raisebox{6.6cm}[0cm][0cm]{(b)}}
        \label{fig:image2}
    \end{minipage}
    \caption{\justifying The SQLS is compared to the VQLS in terms of resource usage. Panel (a) shows the schemes for the different circuit architectures employed by the two algorithms in evaluating a single term in the cost function (i.e., Eq.~\eqref{eq:local_cost_2}) where: (a)(i) is the schematic quantum circuit for the SQLS, (a)(ii) and (a)(iii) are the schematic circuit implementations for the VQLS. Panel (b) shows the scaling of the number of circuits per optimization step (or evaluation of the cost function), $N_{circuits}/step$, comparing between the VQLS and SQLS, respectively. The estimate is provided for a $2^{50} \times 2^{50}$ linear system of the form given in Eq.~\eqref{eq:rqlsp}, in which $k$ indicates the locality of the Pauli strings forming the linear system. Evidently, as the number of terms in the linear system $L$ increases (i.e., when the linear system becomes more complex), the SQLS has a major advantage in resource usage as compared to VQLS.}
    \label{fig:resouce_investigation}
\end{figure*}

\section{Results}
\label{sec:results}
We now report a number of theoretical studies aimed at 
granting an analytic and heuristic understanding of the SQLS. 
Section~\ref{sec:resources} is devoted to understanding the resource usage 
of the SQLS, and comparing it to the VQLS~\cite{VQLS}, which is the currently most 
resource-efficient variational approach to the QLSP, to our knowledge. 
Then, in Sec.~\ref{sec:convergence} we study the heuristics of the 
convergence times of the SQLS for a number of different linear systems. We note that the largest linear system solved is of dimension 4. 
Finally, in Sec.~\ref{sec:Laplace} we are able to leverage techniques
from linear algebra to apply the SQLS to a real physics problem, 
the discretized Laplace equation in a 2D grid, and find its solution. 
In the following, all quantum circuits are simulated using the python Pennylane 
library~\cite{SQLS_dataset,bergholm2022pennylane}.

\subsection{Resource investigation}\label{sec:resources}
The first heuristic study carried out on the SQLS looks at
quantifying the resource used in performing the evaluation
of the cost function defined in Eq.~\eqref{eq:local_cost_2}. 
Within the investigation, we consider three variables: number of qubits, circuit depth and number of circuits 
per cost function evaluation. 
We start with the comparison between the SQLS and the VQLS,
by analysing the requirements in terms of circuit depth and  number of qubits to evaluate a single term in the cost function. 
We will assume that both algorithms have the same 
$n = \log N$ qubit ansatz $V(\Vec{\theta})$. 
The assumptions made on the linear system are the same as in 
Sec.~\ref{sec:sqls}, i.e. the linear system can be expressed as in 
Eqs.~\eqref{eq:paulidecomp_A} and~\eqref{eq:paulidecomp_U}, 
and the cost function is given in  Eq.~\eqref{eq:local_cost_2}.\\
For the SQLS, the scaling of both the circuit depth and the number of qubits is solely determined by the ansatz $V(\Vec{\theta})$. 
In fact, as described in Sec.~\ref{sec:sqls}, at each 
iteration the SQLS creates a classical shadow
$S(\rho_x, N_{shadow})$ that is then used to approximate all
the required expectation values in Eq.~\eqref{eq:local_cost_2}.
The creation of the shadow is done 
by applying $V(\theta)$ to $\ket{0}^{\otimes n}$, 
followed by the application of a random Pauli string with a locality based on the terms in Eqs.~\eqref{eq:paulidecomp_A} 
and~\eqref{eq:paulidecomp_U}, 
before performing a measurement over the whole register. 
Since no additional ancillary qubit is required, 
$n_{SQLS} = \log N$. 
Furthermore, since the application of a Pauli string has depth $1$, the scaling of the circuit depth
is determined by the depth of the ansatz $V(\Vec{\theta})$ plus a constant.
The circuit implementation of the SQLS can be schematically seen in 
Fig.~\ref{fig:resouce_investigation}a(i). \\
For the VQLS, instead, the scaling of the circuit depth is 
determined by the ansatz as well as the locality of the terms in 
Eqs.~\eqref{eq:paulidecomp_A} and~\eqref{eq:paulidecomp_U}; in this case, the minimal number of qubits required is $n_{VQLS} = \log N + 1$. 
This is because the VQLS uses the Hadamard test to estimate the expectation values,
a procedure that exploits an extra ancillary qubit (i.e., the $+1$ in the expression above).
In addition, the Hadamard test requires large controlled operations to calculate the terms
$\delta_{ijlp}^{r}$ (Eq.~\eqref{eq:mu}) and $\beta_{ij}$ (Eq.~\eqref{eq:omega}).
The circuit implementation to calculate 
$\beta_{ij}$ and $\delta_{ijlp}^r$ can be schematically seen in 
Figs.~\ref{fig:resouce_investigation}a(ii) 
and~\ref{fig:resouce_investigation}a(iii), respectively.
Although the controlled unitaries shown in these figures could be merged into a single controlled operation, we keep them separate to clearly show the individual steps involved in the Hadamard-test evaluation.
Full details on the Hadamard test to calculate 
$\beta_{ij}$ and $\delta_{ijlp}^r$ are given in 
App.~\ref{sec:hadtest}, for completeness.
Therefore, the SQLS uses $1$ less qubit and, most importantly, removes the 
need for large controlled unitaries outside the ansatz $V(\Vec{\theta})$, which typically require a large number of two-qubit gates.
We further notice that a classical and computationally cheap pre-processing 
can be performed to reduce both the locality 
and the total number of terms
to calculate, as described in App.~\ref{sec:preprocessing}. 
On the one hand, this would allow for a reduction in the circuit depth of the VQLS, on the other hand it would guarantee a locality reduction of the terms, thus reducing the number of shadows required by Eq.~\eqref{eq:shadowbound}. 
Nevertheless, the Hadamard test would still require large controlled operations, meaning that the SQLS would still be advantageous 
both in terms of the number of qubits and circuit depth. 

The third point of comparison is the number of circuits required to take an optimization step, 
which we denote as $N_{circuits}/step$. 
To simplify the analysis, 
we assume the linear system to be described by the following matrices
\begin{equation} \label{eq:rqlsp}
\begin{split}
A & = \sum\limits_{i=1}^{L} c_i A_i^{k} \\
U & = H^{\otimes n},
\end{split}
\end{equation}
in which $c_i \in [-1,1]$
and $A_i^k$ are $k$-local Pauli strings. This form of linear system
allows to characterize $N_{circuits}/\text{step}$ as a function of a single
parameter, $L$, indicating the number of Pauli strings in the system, 
rather than having two parameters as in Eqs.~\eqref{eq:paulidecomp_A} and~\eqref{eq:paulidecomp_U} ($L_A$ and $L_U$). Furthermore, having all Pauli strings with fixed locality allows 
to investigate the effects of the locality on $N_{circuits}/step$.  
Within the VQLS framework, the number of circuits
(each evaluated with a precision~$\epsilon_{hadamard} = 1/\sqrt{n_{shots}}$)
required to evaluate the cost function of Eq.~\eqref{eq:local_cost_2}
for a linear system expressed as in  Eq.~\eqref{eq:rqlsp} 
is given by:
\begin{equation}
    \label{eq:countvqls}
    \left( \frac{N_{circuits}}{step} \right)_{VQLS} = \frac{n_{shots}}{2} \left( L(L-1) + nL^2 \right),
\end{equation}
where we assume that all circuits are executed for the same number of 
shots, $n_{shots}$. 
Details on the derivation of Eq.~\eqref{eq:countvqls} 
can be found in App.~\ref{sec:circcount}. \\
Within the SQLS, instead,  $N_{circuits}/ step$ is given by Eq.~\eqref{eq:shadowbound}. Using the same considerations as above for the VQLS, the number of circuits per step is given by:
\begin{equation}
    \label{eq:countsqls}
    \left( \frac{N_{circuits}}{step}\right)_{SQLS} = \frac{\log((L(L-1) + nL^2))3^{2k+1}}{\epsilon_{\text{shadow}}^2}.
\end{equation}
Here, the constant of proportionality in Eq.~\eqref{eq:shadowbound} is set to $1$, which our numerical
experiments reported in Sec.~\ref{sec:convergence} show to be a
conservative value.
The power of $2k + 1$ is a worst-case scenario for the terms
$\delta_{ijlp}^r$. 

From a direct comparison of the equations above, we can see that the SQLS has an exponentially better scaling of 
the $N_{circuits}/step$ with respect to the number of terms in the
linear system ($L$) and the number of qubits ($n$). 
However, a drawback of the SQLS that is directly inherited from the classical shadows approach is the exponential scaling in the locality of the Pauli strings, $k$, which is the reason behind our initial low-locality assumption (see Sec.~\ref{sec:sqls}). 
A graphical representation of the scaling of 
$N_{circuits}/step$ as a function of  $L$ is reported for both algorithms in 
Fig.~\ref{fig:resouce_investigation}b. 
These results are obtained by assuming to have $n=50$ qubits, corresponding to a $2^{50} \times 2^{50}$ linear system dimension, and a   
variable number of Pauli strings, $L \in [4, 2500]$, with
locality $k \in [2,3,5,7]$ for the SQLS case. 
The choice for the $L$ domain is 
motivated by the assumption that it scales polynomially with $n$ (i.e., we assumed $L_{max}=n^2=2500$).
To ensure the same error in both procedures, we set
$\epsilon_{hadamard} = \epsilon_{shadow} = 0.01$ for all the points, i.e., for the VQLS
we took $n_{shots}=10,000$.
We stress that Fig.~\ref{fig:resouce_investigation}b reports only the estimated scaling of the required resources, and it does not involve solving an actual $2^{50} \times 2^{50}$ linear system.
The plot in Fig.~\ref{fig:resouce_investigation}b demonstrates the exponential advantage of the SQLS as compared to the 
VQLS in terms $N_{circuits}/step$, for limited locality of the Pauli strings. 
We also stress that the pre-processing technique described in App.~\ref{sec:preprocessing}
can be used to simplify the linear system. 
In fact, it reduces the number and locality
of the expectation values to evaluate the cost function. 
Anyway, this does not
change the favourable scaling of the SQLS. The full investigation
of the $N_{circuits}/step$ with pre-processing can be found in 
App.~\ref{sec:Ncircpp}.

\subsection{Convergence time}\label{sec:convergence}
In the second heuristic study we investigate the convergence times of the SQLS.
In particular, we compare this characteristic timescale to the corresponding one obtained in the
VQLS. In the noiseless limit, the latter has been shown to provide
a scaling as $\text{poly} (\log (N)) \kappa \log(1/\epsilon)$. 
The convergence study was carried out with both methods for $4$ different systems of equations, as follows.

First, we considered a $16 \times 16$ QLSP inspired by the Ising model~\cite{Tacchino_2019} (IQLSP), 
described by the following operations
\begin{equation} \label{eq:ising-qlsp}
\begin{split}
A_{IQLP} & = \frac{1}{\zeta} \left( \sum\limits_{i=1}^{n} \sigma^x_j + J\sum\limits_{i=1}^{n-1} \sigma^z_j \sigma^z_{j+1} + \eta \mathbb{I} \right) \\
b_{IQLSP} & = H^{\otimes n}\ket{0}.
\end{split}
\end{equation}
Then we considered two $16 \times 16$ randomly generated QLSP (also named RQLSP), as 
summarized by these operations 
\begin{equation} \label{eq:random-qlsp}
\begin{split}
A_{RQLSP} & = \frac{1}{\zeta} \left( \sum\limits_{i=1}^{L}  c_i 
    P_i^k + \eta \mathbb{I} \right) \\
b_{RQLSP} & = H^{\otimes n}\ket{0}.
\end{split}
\end{equation}
Full details on these linear systems can be found in App.~\ref{sec:linearsystems}. Furthermore we also study the convergence time of the Potential Grid Linear System (PGLS), introduced in Sec.~\ref{sec:Laplace}.

\begin{figure}
    \centering
    \includegraphics{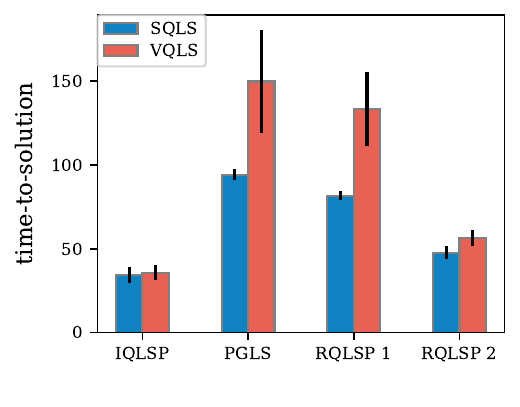}
    \caption{\justifying A comparison of the time-to-solution for a number of different linear systems as obtained numerically by applying either the SQLS or the VQLS, respectively. We plot the average over 10 runs and the error bars are plotted using the standard deviation over the 10 runs.}
    \label{fig:Convergence time}
\end{figure}

The numerical experiments start with the creation of the linear systems of equations. The IQLSP is created with $J=0.1$, and adjusting $\eta$ and $\zeta$ such that the condition number is $\kappa = 60$ (see App.~\ref{sec:linearsystems} for the  explicit values). 
For the RQLSP, the locality of the matrices is set to $k=2$,
and $\eta$ and $\zeta$ are adjusted such that the condition number
for the matrix is $\kappa = 10$ to have manageable convergence 
times. Following these settings, all the examples are subjected to the classical  
pre-processing procedure, described in App.~\ref{sec:preprocessing},
which reduces the complexity 
and quantity of expectation values to be calculated 
and, therefore, reduces the computational times of the
simulations.\\ 
Then, a problem-specific ansatz is chosen for the different linear systems of equations. In the cases of the IQLSP and the PGLS, a real amplitude 
ansatz is used with 4 layers, whilst for the RQLSP a hardware efficient ansatz is used with 10 layers.
Full details on the parametrized circuit designs can be found in App.~\ref{sec:ansatz}. The numerical experiments consisted in solving each linear system of equations with either the SQLS or the VQLS. 
All systems are solved 10 times with different small parameter initializations~\cite{zhang2022escapingbarrenplateaugaussian, Park_2024, wang2023trainabilityenhancementparameterizedquantum}.
In addition, for all experiments
$\epsilon_{hadamard} = \epsilon_{shadow}$, with the following values: $0.01$ for the IQLSP, and $0.03$ for RQLSP1, RQLSP2, and the PGLS (for the VQLS, these correspond to $n_{\mathrm{shots}} = 1/\epsilon_{\mathrm{hadamard}}^{2}$). For the SQLS, the constant of proportionality related to Eq.~\eqref{eq:shadowbound} is set to 1.\\
Following the work in Ref.~\cite{vqlsoptimizer}, we choose the numerical optimizers to be
the modified Powell method~\cite{powell} and  
Adam~\cite{kingma2017adam}, which are shown to 
have similar convergence times. These optimizers are implemented through the
Python Scipy library (Powell)~\cite{2020SciPy-NMeth}, 
and the the Pennylane library
(Adam optimizer)~\cite{bergholm2022pennylane}. 
Due to the simplicity of the IQLSP, which guarantees smaller 
convergence times, the Powell optimizer
is used for this system, whilst the RQLSP1, RQLSP2 and PGLS use
the Adam optimizer, since it guarantees slightly better convergence
times. 
When the Adam optimizer is used, it is initiated with a learning rate of $0.01$, which stays constant throughout the entire optimization process.
The latter is terminated when the trace distance, Eq.~\eqref{eq:error},
reaches a threshold value. 
For the IQLSP and the PGLS the 
threshold trace distance is set to $\epsilon=0.01$,
while
for RQLSP1 it is $\epsilon = 0.1$, and for the RQLSP2 we set $\epsilon = 0.15$. These values were chosen to balance computational cost and reliable convergence across different problem instances. The PGLS converges more rapidly and requires fewer ansatz layers and observables, allowing us to reach a smaller target error. The RQLSP instances, instead, involve deeper circuits, a larger number of observables, and higher computational cost. While RQLSP1 consistently converged to $\epsilon = 0.1$ for all random initializations, RQLSP2 did not reliably converge below $\epsilon = 0.15$ across all seeds, and we therefore adopted this value to ensure consistent convergence.\\
A summary of the results obtained from these numerical experiments can be seen in Fig.~\ref{fig:Convergence time}. The figure shows a comparison of the time-to-solution for a number of different linear systems as obtained numerically by applying either the SQLS or the VQLS, respectively. By time-to-solution we mean the number of cost-function evaluations required to guarantee a solution with precision $\epsilon$ (see Eq.~\eqref{eq:error}). We stress that $\epsilon$ is a measure for the  precision in the solution, whilst $\epsilon_{hadamard}$ and $\epsilon_{shadow}$ are the precision of the techniques in calculating the expectation values.
For all the systems we see that, for equal error, the SQLS has an advantageous/comparable convergence time when compared to the VQLS. 
We point out that this outcome is particularly relevant and noteworthy for two reasons. 
First, it guarantees the SQLS better overall resource usage as compared  
to the VQLS. In fact, given the arguments laid out in Sec.~\ref{sec:resources},
if the SQLS takes less/comparable time to converge and uses fewer resources per
optimization step (in terms of the number of qubits, circuit depth and total 
$N_{circiuts}$ at equal error), then  it has a better resource usage overall.
Second, we claim that the convergence of the SQLS in the noiseless limit is upper bound by
$\mathcal{O}(\text{poly} (\log(N)) \kappa \log(1/\epsilon))$, which is the
convergence of the noiseless VQLS. This guarantees the SQLS a
poly-logarithmic scaling in the system size, which is an exponential advantage when compared to the best
classical algorithm, and makes it a difficult problem to simulate classically.\\
\subsection{Electrostatic Potentials in a Grid}\label{sec:Laplace}
As an example of practical application of the SQLS to a problem of physical nature, we show the results of the solution of 
a linear system derived from the discretization of the classical Laplace equation, i.e., $\nabla^2 \Vec{\phi} = 0$, on a two-dimensional (square) grid.
We present this use case since it represents a linear system derived from the Laplace matrix~\cite{Spectralgraphtheory}, i.e., an object that commonly appears in various fields, such as random walks and analysis of electrical networks~\cite{doyle2000randomwalkselectricnetworks}.

\begin{figure}[t]
    \centering
    \includegraphics[width=1\linewidth]{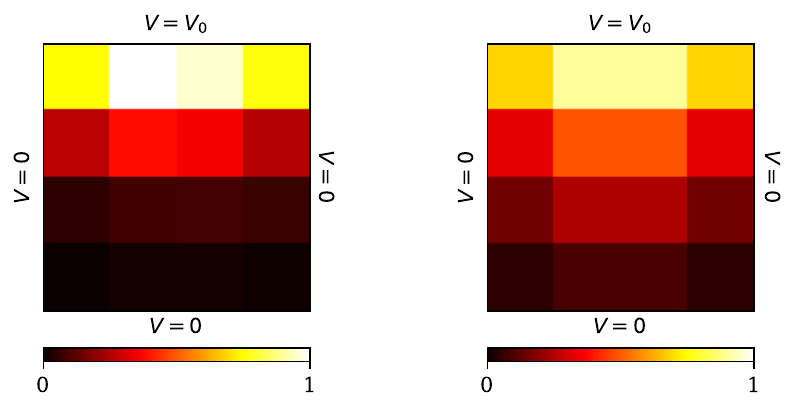}
    \caption{\justifying The potentials derived form the solution of a $16 \times 16$ discretized Laplace equation onto a $4 \times 4$ 2D grid. In (a) we show the solution that was reached using the SQLS, whilst in (b) we show the exact analytic solution. The solution reached with the SQLS has a 0.99 fidelity with respect to the analytic one. Both solutions have been normalized to have the same color range and be directly comparable.}
    \label{fig:potentials}
    \shiftleft{4.5cm}{\raisebox{7.45cm}[0cm][0cm]{(a)}}
    \shiftleft{-.25cm}{\raisebox{7.45cm}[0cm][0cm]{(b)}}
\end{figure}

Essentially, the system we will be looking at is a $\sqrt{N} \times \sqrt{N}$ grid of potentials 
with a constant potential $V_0$ on the top boundary, and a null potential
on all other boundaries. 
In this case the potentials inside the grid, $\Vec{\phi}$,
will be given by the linear system $A\Vec{\phi}=\Vec{b}$, where
$A$ is a $N \times N$ matrix, i.e.,  Eq.~\eqref{eq:matA_potentials}, 
and $\Vec{b}$ is a vector of size $N$, i.e., Eq.~\eqref{eq:b_potentials}.
\begin{equation}\label{eq:matA_potentials}
A_{i,j} = 
  \begin{cases}
        4, & \text{if } i=j\\
        -1, & \text{if } |i-j|=1 \text{ or } |i-j|=\sqrt{N}\\
        0, & \text{Otherwise}
    \end{cases}.
\end{equation}
\begin{equation}\label{eq:b_potentials}
\Vec{b} = 
  \begin{cases}
        V_0, & \text{if } i\leq \sqrt{N}\\
        0, & \text{Otherwise}
    \end{cases}.    
\end{equation}
The detailed derivation of the linear system from the continuous variables Laplace equation can be found in App.~\ref{sec:linsysderivation}.
For the numerical experiments presented here, we take $N=16$, which corresponds
to the use of $n=\log(16)=4$ qubits on a quantum register.
In order to satisfy the norm condition on the matrix $A$,
i.e., $\|A\|_2 \leq 1$, the matrix has to be normalized such that the values on the diagonal are equal to 
$0.22941573$, and the off-diagonal values were equal to $-0.05735393$.
This left the choice of $V_0=0.5 \times ||A||_2$
such that the normalization condition of $\Vec{b}$ is fulfilled after the normalization of the linear system.  Thus the implementation on a quantum computer can be done using the unitary operation $U = \mathbb{I}^{\otimes 2} \otimes H^{\otimes 2}$.\\
With this linear system at hand, the matrix $A$ is decomposed into a linear combination of $4$ unitaries using the theorem presented 
in Ref.~\cite{pedersen1989analysis}. More details on this algebraic result are given in App.~\ref{sec:matdecomp}, for completeness. 
Since the matrix $A$ is sparse, also the decomposed matrices will be 
sparse. Therefore, further decomposition can be achieved by 
implementing the algorithm developed by Ref.~\cite{hantzko2024tensorized} 
to decompose each unitary as a linear sum of Pauli strings.
The combination of these two procedures allows us to transform the full matrix
$A$ of Eq.~\eqref{eq:matA_potentials} into a form that is suitable for the SQLS implementation.
We also notice that other algorithms exist, which allow to decompose a matrix $A$ into linear sums of Pauli strings~\cite{pesce2021h2zixy,Vidal_Romero_2023,bergholm2022pennylane,
jones2024decomposing, hantzko2024tensorized}.
The system is then pre-processed by using the procedure described in App.~\ref{sec:preprocessing}. 
The parameters are initialized to small values, which helps the optimization process~\cite{zhang2022escapingbarrenplateaugaussian, Park_2024, wang2023trainabilityenhancementparameterizedquantum}. 
The Adam optimizer~\cite{kingma2017adam} is chosen for the classical optimization steps, where the starting
learning rate is set to $0.01$. For the optimization process, 
$\epsilon_{shadow} = 0.03$.
This optimization has finally produced an approximate solution with $0.99$ fidelity with the exact analytic solution of  the same problem. The results are explicitly shown on the $4\times 4$ grid in Fig.~\ref{fig:potentials}. In particular, the potential distribution of the approximate solution from the SQLS, shown in Fig.~\ref{fig:potentials}a, is compared to the exact solution of the same linear system of equations, represented on the same grid in Fig.~\ref{fig:potentials}b. 
We stress that the SQLS was able to produce a solution to a high precision, which is a good motivation that it might become a valid approach in solving real world linear systems of practical usefulness.

\section{Discussion\label{sec:dis}}
We have presented the Shadow Quantum Linear Solver (SQLS), 
a resource-efficient quantum 
algorithm to find the solution to the Quantum Linear System Problem (QLSP). 
By building upon recent work on 
variational algorithms to solve linear systems~\cite{VQLS}, and the resource-efficient framework of classical shadows~\cite{Huang_2020}, 
the SQLS appears to be the most promising variational algorithm to solve real systems of equations on NISQ hardware due to its 
convenience in resource usage, convergence times, and resilience to noise. 

In particular, we have presented a series of analytic and heuristic 
studies that quantify various aspects of the SQLS. First, we find that the SQLS has better usage of the number of qubits, 
circuit depth, and number of circuits per cost function evaluation, when
compared to the approach already present in the literature~\cite{VQLS}, which is 
the most resource-efficient variational algorithm to the QLSP known so far for limited locality,
in all the three categories mentioned above. 
Furthermore, we have shown that the convergence times between the two algorithms, 
for an equal error, is more favorable for the SQLS. This is an indication that, in the noiseless limit, the SQLS has a scaling that goes as $\mathcal{O}(\text{poly} (\log(N)) \kappa \log(1/\epsilon))$, in addition to an exponential advantage when compared to the best classical algorithms to solve large linear systems of equations. More than that, the resilience to global depolarization noise offered by the SQLS makes it an ideal candidate for 
current NISQ hardware.
Finally, we were able to leverage techniques from linear algebra to apply
the SQLS to a real physics problem, the discretized Laplace equation on 
a 2D grid. \\
We also notice that the SQLS has room for significant improvement.
On the side of classical shadows, 
further improvements would include the implementation of a number of recent 
works that show how classical shadows can be optimized for the approximation of sums of expectation values of Pauli strings. 
Some of these techniques include the
randomization schemes~\cite{Huang_2021,huang2023learning},
using locally biased classical shadows~\cite{hadfield2020measurements},
principal eigenstate shadows~\cite{grier2024principal}, and dual frame optimization methods~\cite{fischer2024dual}. 
These techniques could either reduce the convergence times, for constant circuit executions, or diminish the number of executed circuits, for constant error.
On side of Variational Quantum Algorithms side, in this work we have applied warm starting to the optimization process 
to solve the discretized 2D Laplace equation. Further studies may 
investigate this technique more in depth~\cite{truger2024warmstarting}.
Other improvements include the possibility to use a 
dynamic ansatz~\cite{Patil_2022}.
An improved SQLS would then be subjected to a heuristic study to investigate the dependence of the convergence time on: 
the condition number $\kappa$, 
the error $\epsilon$, and the system size $N$.

In summary, we believe that the method presented here could accelerate the development of practical solutions to the QLSP on NISQ devices. Furthermore, the paper highlights methods for converting complex linear systems into QLSPs, potentially inspiring novel approaches for
solving systems of linear equations using quantum computers.

\section*{Acknowledgments}
The authors acknowledge useful scientific discussions with A. Abbas, D. Cugini, S. El-Shawa, C. Macchiavello, D. Monaco, A. R. Morgillo, D. Nigro, G. Pellegrini,
P. Perinotti, S. Roncallo, F. Tacchino, and S. Zanotti. 
F. Ghisoni would also like to acknowledge the use of high-performance computing cluster EOS at the Department of Mathematics, University of Pavia. D. Bajoni and D. Gerace acknowledge partial support from the `National Centre For HPC, Big Data and Quantum Computing' (CN1, Spoke 10) within the Italian `Piano Nazionale di Ripresa e Resilienza (PNRR)', Mission 4 Component 2 Investment 1.4 funded by the European Union - NextGenerationEU - project CN00000013.
F. Scala and F. Ghisoni were supported by the `National Quantum Science Technology Institute' (NQSTI, PE4, Spoke 1) within the PNRR  project PE0000023. 

\bibliography{bibliography}

\appendix

\section{Optimal Parameter Resilience}\label{sec:OPR_res}
In Ref.~\cite{VQLS} it was proven that the cost function in Eq.~\eqref{eq:local cost} exhibits OPR to either global depolarization or measurement noise, respectively.
Here we argue that the OPR to depolarization noise is maintained in the SQLS, whilst the OPR to measurement does not carry through. \\
The proof for OPR to global depolarization noise 
was based on 2 assumptions:
the noise affects the estimation of the real and imaginary parts of all
the terms in equal amounts, and the depth of the circuit is dominated by the
ansatz. With regards to the first assumption, since the real and imaginary parts are calculated using the same shadows that are subjected
to the same noise, then both would be equally affected. Concerning the
second assumption, this is especially true in the SQLS, in which the circuit
depth is wholly determined by the ansatz. Since the two assumptions on which the argument for OPR to global
depolarization noise for the VQLS hold also for the SQLS, it is possible to conclude that the SQLS has OPR to a global depolarization 
noise, simply following Ref.~\cite{VQLS}. \\
Regarding OPR to measurement noise, instead, we notice that it cannot be straightforwardly translated to the SQLS, and further studies are required to investigate its actual effects.
First, we will assume that OPR to measurement noise in the context 
of the SQLS refers to OPR to the number of collected shadows. We believe
this is a fair comparison since, just like with shots, the collection
of shadows is what allows for the accuracy of the solution.
We recall that the original proof of OPR to measurement noise
is reliant on the modelling of each single-qubit measurement as a local
depolarization channel. If the circuit measures only one qubit, which is 
the case for the Hadamard test, then the local depolarization channels
can be modelled as a global depolarization channel. Since we already proved
the OPR to a global depolarization channel, then the cost function
also has OPR to measurement noise.
The SQLS does not fit into this proof since the collection of the classical shadows
requires the measurement of all qubits, and then the measurement noise cannot be modelled as a global depolarization noise, which allows to conclude that OPR to measurement noise (number of shadows) does not hold.

\section{Hadamard test}\label{sec:hadtest}
Here we explain the details of the Hadamard test to calculate
the terms $\beta_{ij}$ and $\delta_{ijlp}^{r}$ of 
Eq.~\eqref{eq:local_cost_2}. 
The calculation of $\beta_{ij}$ requires 
$n = \log_2(N)$ qubits, which are required to encode the state $V(\Vec{\theta})\ket{0}^{\otimes n}$. 
An additional qubit is used as ancilla,  initialized in the state $H\ket{0} = \ket{+}$, which is used to sequentially perform $C_aA_i^{k_i}$ and $C_aA_j^{k_j\dagger}$,
where $C_aG$ represents the application of gate $G$ on the 
main register controlled on the ancilla qubit. 
Finally, the application of a Hadamard gate followed by a measurement
on the ancilla qubit is used to calculate the real part of  
$\beta_{ij}$. 
The estimation of the imaginary part of $\beta_{ij}$ requires the ancilla 
to be initialized in the state $S^{\dagger}H\ket{0}$, instead, and then followed by the same operations. The circuit implementation can be seen in 
Fig.~\ref{fig:resouce_investigation}a(ii).

The calculation of $\delta_{ijlp}^{r}$ also requires
$n = \log_2(N)$ qubits, which are needed to encode  the state $V(\Vec{\theta})\ket{0}^{\otimes n}$, as well as an additional ancillary qubit initialized in the state $\ket{+}$.
Here, the operations to perform are:
$C_aA_i^{k_i}$, $C_aU_l^{k_l\dagger}$, $C_a\sigma^z_r$, $C_aU_p^{k_p}$ and $C_aA_j^{k_j\dagger}$, where $C_a\sigma^z_r$ is a $\sigma^z$ on the $r$-th qubit controlled on the ancilla. As before the imaginary part requires starting the ancilla 
qubit in the state $S^{\dagger}H\ket{0}$. 
This circuit is represented in Fig.~\ref{fig:resouce_investigation}a(iii).
We notice that problem specific circuits can be devised to estimate the terms $\delta_{ijlp}^r$ and $\beta_{ij}$. These techniques typically double the number of qubits to reduce the number of large controlled operations to be performed.

\section{Circuit count for the VQLS}\label{sec:circcount}
Here we explain the details of the calculation of $N_{circuits}/step$
for the VQLS protocol, reported as a result in the main text, see  Eq.~\eqref{eq:countvqls}. 
Due to the new form of the linear systems reported in Eq.~\eqref{eq:rqlsp}, 
the calculation of the cost function requires to accurately
estimate the terms 
$\delta_{ij}^{k} = 
\langle x |  A_j^\dagger U \sigma^z_r U^{\dagger} A_i |x\rangle$
and 
$\beta_{ij} = \langle x| A_j^\dagger A_i|x\rangle$.
We will start by considering the terms $\beta_{ij}$. By inspection, we see that if $i=j$ then $\beta_{ii} = 1$. 
Further inspection allows us to determine that   $\beta_{ij} = \left(\beta_{ji} \right)^{*}$, which, combined with the symmetry of the sum, implies that
there is no need to calculate the imaginary part of the expectation values since they will all cancel out.
This leaves $\frac{1}{2}L\times (L-1)$ terms to approximate 
$\omega$ (Equation~\eqref{eq:omega}).

As with regards to the term of the form of $\delta_{ij}^{k}$,
we notice that we can quarter the terms to be calculated since:
$\delta_{ij}^{k} = \left(\delta_{ji}^{k} \right)^{*}$,
and we are only interested in finding the real part of each expectation value.
This leaves $\frac{n}{2} \times L^2$ terms to find
$\mu$(Equation~\eqref{eq:mu}). 
Using these considerations and assuming that each circuit
is sampled with a constant number of shots $n_{shots}$, we finally find that 
the total number of quantum circuits that are needed to evaluate the cost function, including the shots, is given by 
\begin{equation}
    \label{eq:countvqls_app}
    \frac{N_{circuits}}{step} = \frac{n_{shots}}{2} \left( L(L-1) + nL^2 \right).
\end{equation}

\section{Pre-processing}\label{sec:preprocessing}
Here we explain the details of the computationally cheap 
pre-processing technique that can be used to simplify linear systems of equations formulated as in Eqs.~\eqref{eq:paulidecomp_A} and~\eqref{eq:paulidecomp_U}. 
The pre-processing performs the following steps:
(i) it creates a list of all expectation values with associated 
coefficients that have to be calculated to evaluate the cost function;
(ii) it then contracts all Pauli strings;
(iii) it groups coefficients of all the contracted expectation values according to the terms in 
Eq.~\eqref{eq:local_cost_2}; (iv) it finally returns 2 lists, one containing the expectation values that have to be calculated for $\omega$ with the related 
coefficients, and one list for $\mu$. \\
Given that the pre-processing operations will include at most scalar addition and Pauli string contractions, which are cheap since they are performed by using Pauli algebra, we can neglect the pre-processing time with respect to the 
total algorithmic time.
The aim of this pre-processing procedure is to reduce the complexity 
and quantity of expectation values to be ultimately calculated in the SQLS. 

\begin{figure}[t]
    \centering
    \includegraphics{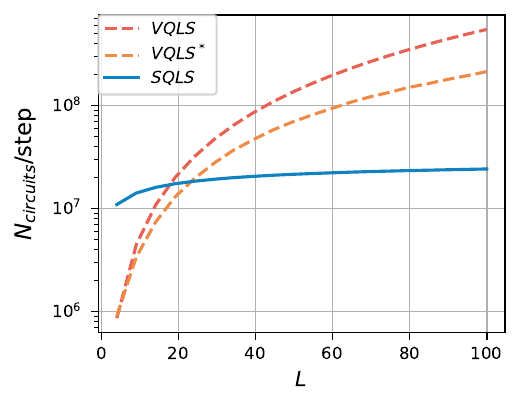}
    \caption{\justifying Results for the quantity $N_{circuits}/step$ as a function of the number of terms in the linear system, $L$, for the different algorithms: VQLS, VQLS with pre-processing ($\text{VQLS}^*$), and the SQLS (also with pre-processing). In this example, the numerical experiment created 30 random $2^{10} \times 2^{10}$ (i.e., 10 qubits dimensional space) linear problems with fixed number of Pauli strings, $L \in [4, 100]$, and fixed locality, $k=2$,  calculating $N_{circuits}/step$ for each procedure. The plot shows that even with pre-processing, the SQLS displays an exponential advantage with the scaling when compared to the VQLS.
}
    \label{fig:resouce_investigation_2}
\end{figure}

\section{Number of circuits per step with pre-processing}\label{sec:Ncircpp}

Here we present the heuristic study for  $N_{circuits}/step$ calculated for the VQLS, VQLS with pre-processing, and the SQLS with pre-processing, respectively. We prefer to present the main scaling result in Fig.~\ref{fig:resouce_investigation} without pre-processing, as it reflects the behaviour of the algorithm for larger and more general linear systems. At the same time, we keep the following analysis to show that the same conclusion holds when pre-processing is applied. In fact, the linear system of equations formalized in  Eq.~\eqref{eq:rqlsp} can also be subjected to a computationally cheap classical algorithm, 
which simplifies number and complexity
of the terms in the cost function, as shown in App.~\ref{sec:preprocessing}. 
A pre-processed cost function can be evaluated using both the Hadamard test and classical shadows.
By defining the number of terms to evaluate the cost function after the pre-processing as $N_{PP}$, the Hadamard test requires $N_{PP} \times n_{shots}$ circuit executions, and it will be referred to as $\text{VQLS}^*$, 
whilst in classical shadows  $N_{circuits}/step$ is determined by Eq.~\eqref{eq:shadowbound}, in which we take $M=N_{PP}$; we will refer to this case as SQLS in the plots.

In our numerical experiment, this study required the creation of 30 random $2^{10} \times 2^{10}$ (10 qubit)
linear problems with fixed
number of Pauli strings, $L \in [4, 100]$, and fixed locality $k=2$.
These were then used to find $N_{circuits}/step$ for: 
the VQLS (Equation~\eqref{eq:countvqls}), 
the VQLS with the pre-processing, $N_{PP} \times n_{shots}$, 
and the SQLS, where the number of operators was $M=N_{PP}$. 
The results for $N_{circuits}/step$ as a function of the number of terms in the linear system, $L$, is reported in Fig.~\ref{fig:resouce_investigation_2}. 
Overall, we see that the scaling of the number of circuits 
per optimization step remains favourable to the SQLS, also 
in the case of pre-processing in the VQLS, which shows a resource
advantage for the SQLS in terms of $N_{circuits}/step$.

\section{Linear Systems}
\label{sec:linearsystems}
Here we report further details about the linear system of equations that were used as part of the comparative investigation reported in Sec.~\ref{sec:convergence}. 
The matrix $A$ for the QLSP inspired by the Ising model (IQLSP) 
is explicitly given by
\begin{multline}
\label{eq:iqlsp_full}
A_{IQLSP} = 0.0123(ZZII - IZZI - IIZZ) + \\
0.123(XIII + IXII + IIXI + IIIX) + 0.508IIII    
\end{multline}
The corresponding matrices for the Random QLSP (RQLSP) are explicitly given by
\begin{multline}\label{eq:rqlsp_full_1}
A_{RQLSP1} = -0.0513IXXI - 0.366IIYY - 0.0352XXII\\
+ 0.144IXIZ + 0.55IIII    
\end{multline}
and 
\begin{multline}\label{eq:rqlsp_full_2}
A_{RQLSP2} = 0.242ZZII -0.0817IZZI  + 0.183XIIX\\
- 0.0780IZIY + 0.55IIII \, ,
\end{multline}
respectively. 
\section{Variational ansatz}\label{sec:ansatz}

\begin{figure}[ht] 
    \centering
    \begin{minipage}[t]{1\textwidth}
    \centering
          \Qcircuit @C=0.5em @R=0.75em {
    & \qw & \gate{R_X(\theta_1)} & \gate{R_Y(\theta_5)} & \gate{R_Z(\theta_9)}    & \ctrl{1} & \qw       & \qw      & \targ & \qw  \\
    & \qw & \gate{R_X(\theta_2)} & \gate{R_Y(\theta_6)} & \gate{R_Z(\theta_{10})} & \targ    &  \ctrl{1} & \qw      & \qw   & \qw    \\
    & \qw & \gate{R_X(\theta_3)} & \gate{R_Y(\theta_7)} & \gate{R_Z(\theta_{11})} & \qw      & \targ     & \ctrl{1} & \qw   & \qw    \\
    & \qw & \gate{R_X(\theta_4)} & \gate{R_Y(\theta_8)} & \gate{R_Z(\theta_{12})} & \qw      & \qw       & \targ    & \ctrl{-3} & \qw
            }
    \vspace{15pt}
    \end{minipage}
    
    \begin{minipage}[t]{1\textwidth}
        \centering
        \Qcircuit @C=0.25em @R=0.75em {
    & \qw & \ctrl{1} & \gate{R_Y(\theta_1)} & \qw      & \qw                   & \qw  \\
    & \qw & \targ    & \gate{R_Y(\theta_2)} & \ctrl{1} & \gate{R_Y(\theta_5)}  & \qw  \\
    & \qw & \ctrl{1} & \gate{R_Y(\theta_3)} & \targ    & \gate{R_Y(\theta_6)}  & \qw  \\
    & \qw & \targ    & \gate{R_Y(\theta_4)} & \qw      & \qw                    & \qw
            }
    \end{minipage}

    \begin{minipage}[t]{1\textwidth}
        \shiftleft{8.2cm}{\raisebox{5.75cm}[0cm][0cm]{(a)}}
        \shiftleft{8.2cm}{\raisebox{2.5cm}[0cm][0cm]{(b)}}
    \end{minipage}
    \caption{\justifying (a) The hardware efficient ansatz that was used to find a solution to systems~\eqref{eq:rqlsp_full_1} and ~\eqref{eq:rqlsp_full_2} and (b) the real amplitude ansatz that was used to find the solution to linear systems~\eqref{eq:iqlsp_full} and~\eqref{eq:matA_potentials}}
    \label{fig:ansatz_diag}
\end{figure}
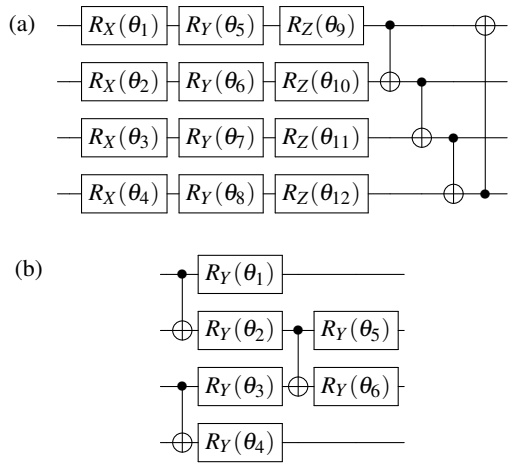

Here we present the ansatzes that were used throughout the numerical experiments whose results are reported in  Sec.~\ref{sec:convergence}.
For the two RQLSP's we used a hardware efficient ansatz of the form represented in 
Fig.~\ref{fig:ansatz_diag}a. Given that these systems were hardware efficient, and they could have involved real and complex elements, the choice of 
this ansatz was to keep the parameter search space as hardware efficient as 
possible. Therefore, all Pauli rotations ($R_X$, $R_Y$, $R_Z$) were used, and CNOTs in a ring structure were used to connect all the elements. 
When applied to a 4 qubits register, this ansatz requires $12$ parameters per layer. 

For the IQLSP and the potential grid, we used the real amplitude 
ansatz, which is reported in Fig.~\ref{fig:ansatz_diag}b. 
The motivation behind this choice was that for both of these linear systems, $A$ and $b$ were only composed of real elements, and therefore also the solution must have been composed of real elements. Therefore, the ansatz that was used was 
only composed of $R_Y$ rotations and CNOTs. 

\section{Matrix Decomposition}\label{sec:matdecomp}
Here we present a simplified version of the
theorem reported in Ref.~\cite{pedersen1989analysis}, which establishes the rule for the decomposition of a matrix into a linear sum of unitary matrices.

First, we notice that any complex matrix can be represented as:
\begin{equation}
  A = B+iC \, ,  
\end{equation}
in which 
\begin{equation}
  B = \frac{1}{2}(A+A^*)  
\end{equation}
\begin{equation}
  C = \frac{1}{2i}(A-A^*)  \, .
\end{equation}
If we assume that $\|A\|_2 \leq 1$, then $\|B\|_2 \leq 1$ and 
$\|C\|_2 \leq 1$ are both true. Therefore, we can further decompose $B$ as:
\begin{equation}
    B = \frac{1}{2}(U_B + V_B) \, 
\end{equation}
with
\begin{align}
    \label{eq:matdecomp1}
    U_B = B + i\sqrt{I-B^2}
    \\ V_B = B - i\sqrt{I-B^2} \, ,
\end{align}
and similarly 
\begin{equation}
    C = \frac{1}{2}(U_C + V_C)   \, , 
\end{equation}
with 
\begin{align}
    \label{eq:matdecomp2}
    U_C = C + i\sqrt{I-C^2}
    \\ V_C = C - i\sqrt{I-C^2} \, .
\end{align}
By inspection, this breakdown can be seen to be true. Furthermore, we can see
that $\left( U_B \right)^* = V_B$ and $\left( U_C \right)^* = V_C$. 
Finally $U_B V_B = V_B U_B = \mathbb{I}$, and 
$U_C V_C = V_C U_C = \mathbb{I}$. In the end, we have decomposed the matrix 
$A$ into a linear combination of 4 unitary matrices of the form
\begin{equation}
    A = \frac{1}{2}U_B + \frac{1}{2}V_B + \frac{i}{2}U_C + \frac{i}{2}V_C \, .
\end{equation}

\section{Local Cost function derivation}\label{sec:costfuncderivation}
Here we explicitly show how equations~\eqref{eq:local_cost_2}~\eqref{eq:mu}~\eqref{eq:omega} are derived from equation~\eqref{eq:local cost}. We will be assuming a QLSP problem of the form $A\ket{x}=\ket{b}$, where we will take the following decompositions:

\begin{align} 
A & = \sum\limits_{i=1}^{L_A} c_i^A A_i^{k_i} \\
U & = \sum\limits_{j=1}^{L_U} c_j^U U_j^{k_j}.
\end{align}

We will start by looking at the equation for the cost function:

\begin{equation}
    C_L(\Vec{\theta}) = \frac{\bra{x(\Vec{\theta})}H_L\ket{x(\Vec{\theta})}}{\braket{\psi(\Vec{\theta})}{ \psi(\Vec{\theta})}}=
    \frac{\bra{x}H_L\ket{x}}{\braket{\psi}{ \psi}}, 
\end{equation}

where $\ket{x}$ is the approximate solution, $\ket{\psi}=A\ket{x}$ and:

\begin{equation}
    H_L = A^{\dagger} U \left( \mathbb{I} - \frac{1}{n}\sum\limits_{r=1}^n \dyad{0_r}{0_r} \otimes \mathbb{I}_r-\right) U^{\dagger}A\, .
\end{equation}

We will first be looking at the term in the denominator, $\bra{\psi} \ket{\psi}$. Here we have that:

\begin{align}
    \bra{\psi} \ket{\psi}  &  = \bra{x} A^{\dagger} A \ket{x} \\
    & = \bra{x} \left( \sum\limits_{j=1}^{L_A} c_j^A A_j^{k_j} \right)^{\dagger} \left( \sum\limits_{i=1}^{L_A} c_i^A A_i^{k_i} \right) \ket{x} \\
    &  = \sum\limits_{i,j=1}^{L_A}c^A_ic_{j}^{A*} \beta_{ij} = \omega. 
\end{align}

Where we have that $\beta_{ij} = \langle x| A_j^{k_j\dagger} A_i^{k_i}|x\rangle$. We now look at the term in the numerator, $\bra{x}H_L\ket{x}$:

\begin{align}
    \bra{x}H_L\ket{x}  & = \bra{x}A^{\dagger} U \left( \mathbb{I} - \frac{1}{n}\sum\limits_{r=1}^n \dyad{0_r}{0_r} \otimes \mathbb{I}_r-\right) U^{\dagger}A\ket{x} \\
    & = \bra{x}A^{\dagger} U U^{\dagger}A\ket{x} - \\ 
    & \bra{x}A^{\dagger} U \left( \frac{1}{n}\sum\limits_{r=1}^n \dyad{0_r}{0_r} \otimes \mathbb{I}_r-\right) U^{\dagger}A\ket{x}.
\end{align}

We can see that the first term, $\bra{x}A^{\dagger} U U^{\dagger}A\ket{x} = \bra{x}A^{\dagger} A\ket{x} = \bra{\psi}\ket{\psi}$. We will now focus on the second term:

\begin{align} 
    & \bra{x}A^{\dagger} U \left( \frac{1}{n}\sum\limits_{r=1}^n \dyad{0_r}{0_r} \otimes \mathbb{I}_r-\right) U^{\dagger}A\ket{x}  \\
    & = \bra{x} \left( \frac{1}{n}\sum\limits_{r=1}^n A^{\dagger} U \left( \dyad{0_r}{0_r} \otimes \mathbb{I}_r- \right) U^{\dagger}A \right) \ket{x} \\
    & = \bra{x} \left( \frac{1}{n}\sum\limits_{r=1}^n A^{\dagger} U \left( \frac{\mathbb{I}_r + \sigma_r^z}{2} \otimes \mathbb{I}_r-\right) U^{\dagger}A \right) \ket{x}
\end{align}

Where in the last line we have used the fact that $\dyad{0}{0} = \frac{\mathbb{I} + \sigma^z}{2}$. Here we get two further terms which, for clarity, we will separate. Firstly we look at the term with the identity:

\begin{align} 
    & = \bra{x} \left( \frac{1}{n}\sum\limits_{r=1}^n A^{\dagger} U \left( \frac{\mathbb{I}_r}{2} \otimes \mathbb{I}_r-\right) U^{\dagger}A \right) \ket{x} \\
    & = \frac{1}{2}\bra{x} \left( \frac{1}{n}\sum\limits_{r=1}^n A^{\dagger} U \mathbb{I} U^{\dagger}A \right) \ket{x} \\
    & = \frac{1}{2}\bra{x} \frac{1}{n}\sum\limits_{r=1}^n \bra{x} A^{\dagger} A  \ket{x}  \\ 
    & = \frac{1}{2} \frac{1}{n}\sum\limits_{r=1}^n \bra{\psi}  \ket{\psi}  \\
    & = \frac{1}{2} \bra{\psi}  \ket{\psi} \\
\end{align}

And for the second term with the $\sigma^z$ we have:

\begin{align}
    & = \bra{x} \left( \frac{1}{n}\sum\limits_{r=1}^n A^{\dagger} U \left( \frac{\sigma^z_r}{2} \otimes \mathbb{I}_r-\right) U^{\dagger}A \right) \ket{x} \\
    & = \frac{1}{2n}\sum\limits_{r=1}^n \bra{x} A^{\dagger} U \left( \sigma^z_r \otimes \mathbb{I}_r-\right) U^{\dagger}A \ket{x} \\
    & = \frac{1}{2n}\sum\limits_{r=1}^{n} 
    \sum\limits_{i,j=1}^{L_A} \sum\limits_{l,p=1}^{L_U}c^A_ic_{j}^{A*}c^U_pc_{l}^{U*} \delta_{ijlp}^{r} = \mu. 
\end{align}

where $\delta_{ijlp}^{r} = \langle x | A_j^{k_j\dagger} U_p^{k_p} (\sigma^z_r \otimes \mathbb{I}_r-) U^{k_l\dagger}_l A^{k_i}_i |x\rangle$. So putting everything together we have that:

\begin{align}
    C_L(\Vec{\theta}) & =  \frac{\bra{x}H_L\ket{x}}{\braket{\psi}{\psi}}  \\
    &  =  \frac{\braket{\psi}{\psi} -  \frac{1}{2}\braket{\psi}{\psi} -  \frac{1}{2n}\sum\limits_{r=1}^{n} 
    \sum\limits_{i,j=1}^{L_A} \sum\limits_{l,p=1}^{L_U}c^A_ic_{j}^{A*}c^U_pc_{l}^{U*} \delta_{ijlp}^{r}}{\braket{\psi}{\psi}} \\
    & = \frac{1}{2} - \frac{1}{2n} \frac{\mu}{\omega}
\end{align}

in which:
\begin{align}
\mu & = \sum\limits_{r=1}^{n} 
    \sum\limits_{i,j=1}^{L_A} \sum\limits_{l,p=1}^{L_U}c^A_ic_{j}^{A*}c^U_pc_{l}^{U*} \delta_{ijlp}^{r}  \\
\omega & = \sum\limits_{i,j=1}^{L_A}c^A_ic_{j}^{A*}\beta_{ij} \, ,
\end{align}
with  $\delta_{ijlp}^{r} = \langle x | A_j^{k_j\dagger} U_p^{k_p} \sigma^z_r U^{k_l\dagger}_l A^{k_i}_i |x\rangle$
and~$\beta_{ij} = \langle x| A_j^{k_j\dagger} A_i^{k_i}|x\rangle$.

\section{Potential Grid Linear System Derivation}\label{sec:linsysderivation}

We consider the two-dimensional Laplace equation
\begin{equation}
    \nabla^2 \Vec{\phi} = 0
\end{equation}

With the following boundary conditions:

\begin{equation}
    \Vec{\phi}=
    \begin{cases}
    V_0, & \text{on the top boundary},\\
    0,   & \text{on the left, right, and bottom boundaries}.
    \end{cases}
\end{equation}

To obtain a linear system suitable for numerical solutions, we discretize the domain using a uniform grid with spacing $h$ in both directions. Let $\phi_{i,j}$ denote the approximation of $\phi$ at the grid point $(i,j)$.  We collect the values of $\phi_{i,j}$ at interior grid points into a vector $\Vec{\phi} \in \mathbb{R}^N$ (where $N$ is the number of interior points). Boundary values are fixed by the conditions above and are not included in the unknown vector.

At any interior point $(i,j)$, the standard five-point stencil gives

\begin{equation}
    \nabla^2 \Vec{\phi}\approx
    \frac{\phi_{i+1,j}+\phi_{i-1,j}+\phi_{i,j+1}+\phi_{i,j-1}-4\phi_{i,j}}{h^2}.  
\end{equation}

Imposing $\nabla^2\phi=0$ yields

\begin{equation}
    4\phi_{i,j}-\phi_{i+1,j}-\phi_{i-1,j}-\phi_{i,j+1}-\phi_{i,j-1}=0.  
\end{equation}

If a neighbour of $(i,j)$ lies on the boundary, its value is known and can be moved to the right-hand side. We can write the equation as

\begin{equation}
    4\phi_{i,j}-\phi_{i+1,j}-\phi_{i-1,j}-\phi_{i,j-1}=V_0.   
\end{equation}

Collecting the equations for all interior points yields a sparse linear system
\begin{equation}
    A\Vec{\phi}=b,  
\end{equation}

where $A$ is the discrete Laplace matrix (with $4$ on the diagonal and $-1$ for interior nearest-neighbour couplings), and the vector $b$ encodes the Dirichlet boundary values. With the top boundary fixed to $V_0$ and the other boundaries set to $0$, the entries of $b$ are nonzero only for interior points adjacent to the top boundary, for which $b$ contains a contribution of $V_0$.

\end{document}